\documentclass[aps,pra,reprint,superscriptaddress,amsfonts,amssymb,amsmath]{revtex4-1}

\usepackage{graphicx}
\usepackage{dcolumn}
\usepackage{bm}
\usepackage[mathlines]{lineno}
\usepackage{siunitx}
\usepackage{booktabs}
\usepackage{color}
\usepackage{multirow}
\usepackage{romannum}

\begin{document}


\title{Defect charging and resonant levels in half-Heusler Nb$_{1-x}$Ti$_x$FeSb}

\author{Yefan Tian}
\affiliation{Department of Physics and Astronomy, Texas A\&M University, College Station, TX 77843, USA}
\author{Farit G. Vagizov}
\affiliation{Department of Physics and Astronomy, Texas A\&M University, College Station, TX 77843, USA}
\affiliation{Kazan Federal University, 18 Kremlyovskaya Street, Kazan 420008, Russia}
\author{Nader Ghassemi}
\affiliation{Department of Physics and Astronomy, Texas A\&M University, College Station, TX 77843, USA}
\author{Wuyang Ren}
\affiliation{Department of Physics, University of Houston, Houston, TX 77204, USA}
\affiliation{Institute of Fundamental and Frontier Sciences, University of Electronic Science and Technology of China, Chengdu 610054, China}
\author{Hangtian Zhu}
\affiliation{Department of Physics, University of Houston, Houston, TX 77204, USA}
\author{Zhiming Wang}
\affiliation{Institute of Fundamental and Frontier Sciences, University of Electronic Science and Technology of China, Chengdu 610054, China}
\author{Zhifeng Ren}
\affiliation{Department of Physics, University of Houston, Houston, TX 77204, USA}
\author{Joseph H. Ross, Jr.}
\email{jhross@tamu.edu}
\affiliation{Department of Physics and Astronomy, Texas A\&M University, College Station, TX 77843, USA}

\date{\today}

\begin{abstract}
We report $^{93}$Nb and $^{121}$Sb NMR and $^{57}$Fe M\"{o}ssbauer studies combined with DFT calculations of Nb$_{1-x}$Ti$_x$FeSb ($0\leqslant x \leqslant0.3$), one of the most promising thermoelectric systems for applications above 1000 K. These studies provide local information about defects and electronic configurations in these heavily $p$-type materials. The NMR spin-lattice relaxation rate provides a measure of states within the valence band. With increasing $x$, changes of relaxation rate vs carrier concentration for different substitution fractions indicate the importance of resonant levels which do not contribute to charge transport. The local paramagnetic susceptibility is significantly larger than expected based on DFT calculations, which we discuss in terms of an enhancement of the susceptibility due to a Coulomb enhancement mechanism. The M\"{o}ssbauer spectra of Ti-substituted samples show small departures from a binomial distribution of substituted atoms, while for unsubstituted $p$-type NbFeSb, the amplitude of a M\"{o}ssbauer satellite peak increases vs temperature, a measure of the $T$-dependent charging of a population of defects residing about 30 meV above the valence band edge, indicative of an impurity band at this location.
\end{abstract}

\maketitle

\section{Introduction}

\begin{figure}
\includegraphics[width=0.6\columnwidth]{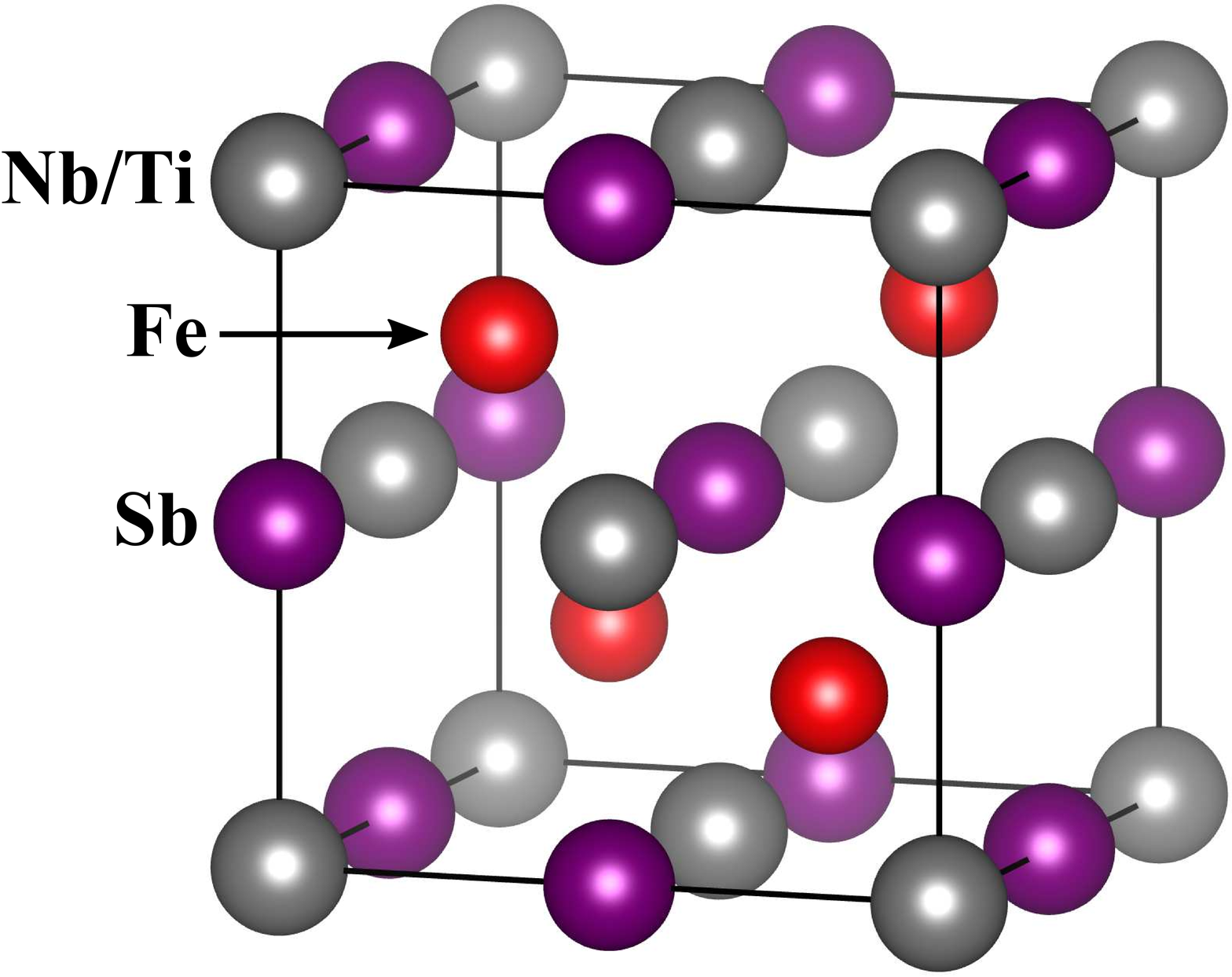}
\caption{\label{Nb1-xTixFeSb_structure}Crystal structure of Ti-substituted half-Heusler Nb$_{1-x}$Ti$_x$FeSb, showing site occupations.}
\end{figure}

The half-Heusler family, as one of the most fascinating intermetallic systems, has gained considerable attention in recent years due to their extraordinary thermoelectric performance and unconventional topological properties. Half-Heusler materials have a general formula $XYZ$ ($X$ representing a (III-V)$_B$ element, $Y$ a transition metal of group VIII$_B$, and $Z$ a tetrel or pnictogen element) \cite{kouacou1995semiconductor,pierre1997properties,tobola1998crossover}. The crystal structure (Fig.~\ref{Nb1-xTixFeSb_structure}) can be formally derived from the cubic Heusler phases $XY_2Z$ by removing one of the two equivalent $Y$ atoms, leaving a structural vacancy. The half-Heusler ideal valence electron concentration is 8 or 18 per formula unit \cite{aliev1989gap,aliev1990narrow,young2000thermoelectric,kandpal2006covalent,graf2011simple}, with semiconducting or semimetallic behavior often observed with those having 18 electrons \cite{pierre1997properties,tobola1998crossover,aliev1989gap,aliev1990narrow,hohl1999efficient}. The rich combination of chemical elements fulfilling this condition leads to a number of interesting properties, from nonmagnetic semiconductors to ferromagnetic half metals, as well as strongly correlated electrons and topological insulator behavior \cite{shi2015nmr,nowak2014nmr,zhang2016nmr}.

Within the half-Heusler system, NbFeSb has recently been of particular interest due to its excellent thermoelectric performance for high-temperature applications and its composition, placing it within the realm of earth abundant thermoelectric materials \cite{he2016achieving}. With various elements substituted, NbFeSb-based semiconductors can exhibit a large power factor, above 100 $\mu$W\,cm$^{-1}$\,K$^{-2}$ \cite{he2016achieving,ren2018ultrahigh,yu2018unique}. Substituted elements can also control the transport properties, thus enabling the electronic behavior to be tuned. With multiple elemental substitutions, a number of different types of half-Heusler alloys have been designed and shown to have high figures of merit ($ZT>1.5$) \cite{fu2015realizing,rogl2017v,zhu2018discovery,zhu2019discovery}. Based on these alloys, thermoelectric generators thus have the potential to reach a high power-conversion efficiency \cite{chen2016half,zhu2018discovery,zhu2019discovery}, enhancing the prospect of future thermoelectric applications. However, native defects can also counteract the desired effects or otherwise degrade the electronic response. The nature of these defects has been explored experimentally for several compounds, such as ZrNiSn and TiCoSb \cite{uher1999transport,wambach2016unraveling,snyder2008complex}. To analyze the underlying electronic and magnetic properties of Nb$_{1-x}$Ti$_x$FeSb, we have performed $^{93}$Nb and $^{121}$Sb NMR and $^{57}$Fe M\"{o}ssbauer measurements as local probes for all sites aiming at a better understanding of these materials.

\begin{table*}
\caption{\label{table1}Substitution fraction ($x$), sample designation (label), actual chemical composition (from microprobe analysis), measured room-temperature carrier concentration ($p$) from Hall measurements and theoretical carrier concentration ($p_\mathrm{theo}$).}
{\footnotesize
\begin{tabular}{>{\centering}p{1.8cm}>{\centering}p{2.5cm}>{\centering}p{3.5cm}>{\centering}p{2.5cm}>{\centering}p{2.5cm}>{\centering}p{1.8cm}}
\hline\hline
\addlinespace[1ex]$x$&label&Actual composition&$p$ ($10^{20}$ cm$^{-3}$)&$p_\mathrm{theo}$ ($10^{20}$ cm$^{-3}$)&$p/p_\mathrm{theo}$\tabularnewline[1ex] \hline
\addlinespace[1ex]0&NbFeSb-1050&NbFeSb\footnotemark[1]&0.9\footnotemark[1]&-&-\tabularnewline[0.5ex]
\addlinespace[1ex]0.05&Ti(0.05)&Nb$_{0.94}$Ti$_{0.05}$Fe$_{1.01}$Sb$_{0.99}$\footnotemark[2]&8.1\footnotemark[2]&9.5&0.85\tabularnewline[0.5ex]
\addlinespace[1ex]0.1&Ti(0.1)&Nb$_{0.89}$Ti$_{0.1}$Fe$_{1.00}$Sb$_{0.99}$\footnotemark[2]&15.2\footnotemark[2]&19&0.80\tabularnewline[0.5ex]
\addlinespace[1ex]0.2&Ti(0.2)&Nb$_{0.8}$Ti$_{0.2}$Fe$_{1.02}$Sb$_{0.99}$\footnotemark[2]&25.7\footnotemark[2]&38&0.68\tabularnewline[0.5ex]
\addlinespace[1ex]0.3&Ti(0.3)&Nb$_{0.69}$Ti$_{0.3}$Fe$_{1.02}$Sb$_{0.98}$\footnotemark[2]&30.3\footnotemark[2]&57&0.40\tabularnewline[0.5ex]
\hline\hline
\end{tabular}}
\footnotetext[1]{From Ref.~\cite{tian2018native}.}
\footnotetext[2]{From Ref.~\cite{he2016achieving}.}
\end{table*}

\section{Experimental and computational methods}

Raw elements (Nb pieces, 99.9\%, and Sb broken rods, 99.9\%, Atlantic Metals \& Alloy; Fe granules, 99.98\%, and Ti foams, 99.9\%, Alfa Aesar) were weighed stoichiometrically, and arc melted multiple times to form uniform ingots. The ingots were then ball milled (SPEX 8000M Mixer/Mill) for 3 h under Ar protection to produce nanopowders. The powders were then consolidated into disks via hot pressing at 80 MPa for 2 min at 1373 K. This process has been shown to yield uniform samples with high power factors \cite{he2016achieving}. In this work, we denote Nb$_{1-x}$Ti$_x$FeSb as Ti($x$) ($x = 0.05, 0.1, 0.2, 0.3$), the same samples as prepared in Ref.~\cite{he2016achieving}. We also studied an unsubstituted sample, which is one of the samples described previously in Ref.~\cite{tian2018native}, annealed at 1323 K (sample NbFeSb-1050).

Substitution fractions, actual compositions and carrier concentrations of all samples are listed in Table~\ref{table1}. Room-temperature carrier concentrations were determined by Hall measurements \cite{he2016achieving} and shown to be $p$-type. Half-Heusler materials normally follow an 18-electron stability rule, and NbFeSb satisfies this criterion and is found to be a semiconductor. Cation substitution in the range (Nb$_{1-x}$Ti$_x$) leads to heavily $p$-type samples because Ti lacks one electron compared with Nb. In these samples, as expected, larger $x$ produces higher hole concentration, however the ratio of the measured charge density to the theoretical charge density becomes smaller with larger $x$ as further discussed below. 

Magnetic measurements were performed using a Quantum Design MPMS superconducting quantum interference device magnetometer. $^{93}$Nb and $^{121}$Sb NMR experiments were carried out by applying a custom-built pulse spectrometer at a fixed magnetic field 9 T using shift standards NbCl$_5$ and KSbF$_6$ in acetonitrile respectively, with positive shifts here denoting paramagnetic sign. Shift is calculated by $\Delta\nu/\nu_0$ where $\Delta\nu=\nu-\nu_0$ is the deviation from the standard reference frequency, $\nu_0$, determined by the shift standard. M\"{o}ssbauer spectra were measured in the temperature range 7-323 K on a conventional constant acceleration spectrometer (WissEl) equipped with a Co-57 source in rhodium matrix 35 mCi in activity. For low-temperature measurements, samples were prepared from fine powders with the density of 18 mg of Fe per cm$^2$ uniformly distributed as a thin layer over sample holder. The samples were mounted on a cold finger of a helium continuous-flow cryostat (CFICEV-MOSS, ICE Oxford, UK), with temperatures controlled within $\pm$0.5 K over the whole temperature range. Above room temperature, a different sample holder was used with a somewhat smaller powder density. Isomer shifts were referred to $\alpha$-Fe at room temperature.

Density functional theory (DFT) calculations were performed with WIEN2k \cite{blaha2001wien2k} using the Perdew, Burke, and Ernzerhof (PBE) exchange-correlation potential, a $k$-point grid of $10 \times 10 \times 10$, and lattice constants from experimental values \cite{tian2018native}. In calculations not including spin-orbit coupling, a semiconducting gap of 0.54 eV was obtained for NbFeSb. This can be compared to 0.51 eV obtained from high-temperature transport measurements \cite{he2016achieving}. For the VB maximum at the $L$ point, we also obtained an effective mass $m_\mathrm{eff} = 4.9\,m_e$ by fitting the calculated density of states within 0.1 eV of the band edge. For $^{93}$Nb NMR chemical shifts the zero offset was calibrated by a separate shift calculation for LaNbO$_4$ and for YNbO$_4$, and then adjusted based on the previously reported shifts \cite{papulovskiy2013theoretical} to the standard reference (NbCl$_5$ in acetonitrile). Since chemical shift are less well studied for $^{121}$Sb, we did not find a comparable solid compound with which to calibrate the computational $^{121}$Sb zero offset. We also performed similar calculations with spin-orbit coupling included, giving relative little change in the results -- for example, the effective mass increases from 4.9 to 5.0\,$m_e$, while the calculated $^{93}$Nb chemical shifts changed by less than 60 ppm. 

\section{Experimental results}

\subsection{NMR measurements}

\subsubsection{Line shapes}

\begin{figure}
\includegraphics[width=0.8\columnwidth]{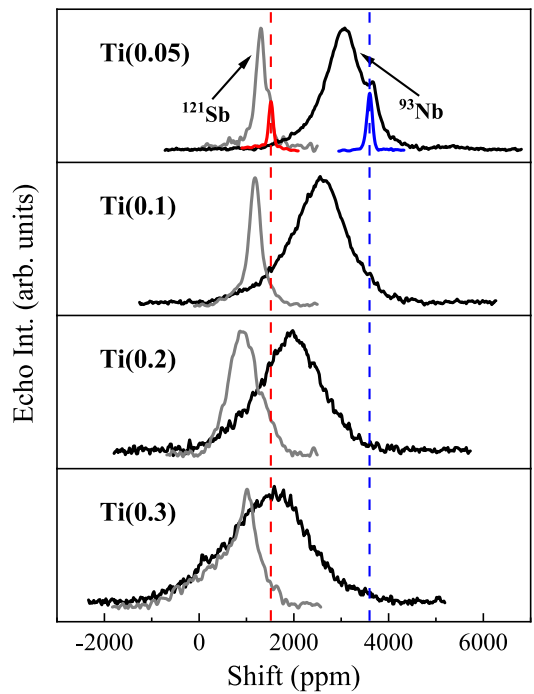}
\caption{\label{Nb1-xTixFeSb_all_rt_lineshape}Room-temperature $^{93}$Nb (black) and $^{121}$Sb (gray) NMR line shapes for Ti($x$) ($x = 0.05, 0.1, 0.2, 0.3$) samples. The previously reported $^{93}$Nb spectrum of NbFeSb (sample NbFeSb-1050) \cite{tian2018native} is shown as a solid curve for comparison, and $^{121}$Sb spectrum is also shown for the same sample, with dashed lines indicating the center positions.}
\end{figure}

Fig.~\ref{Nb1-xTixFeSb_all_rt_lineshape} shows $^{93}$Nb and $^{121}$Sb NMR line shapes for the Ti($x$) samples obtained from the fast Fourier transform using a standard spin echo sequence. Also superposed are resonances for the $x=0$ sample NbFeSb-1050 with a much smaller room-temperature carrier concentration ($p=9\times10^{19}$ cm$^{-3}$) \cite{tian2018native}. It can be seen from the superposed spectra that there is a small signal due to pure-phase NbFeSb, appearing only in the Ti(0.05) substituted sample. The resonances for both nuclei become broader and move to lower frequencies when the substitution fraction increases.

The increasing line width vs $x$ is due to a superposition of local environments at Nb and Sb sites. For Nb, the first and second neighbor shells are composed of Fe and Sb ions respectively, so it is only starting with the third shell that the (Nb, Ti) substitution produces a distribution of local environments. Correspondingly, for Sb the first neighbor shell consisting of Fe ions is fixed, while the second neighbors include a distribution of (Nb, Ti) ions. This leads to relatively symmetric local environments, with a large number of (Nb, Ti) configurations contributing to the observed line widths. Since the widths scale nearly proportionally to the changes in shift relative to NbFeSb, with Sb widths and shifts considerably smaller, we surmise that the widths, as well as the asymmetry seen for Ti(0.3), are due to a quasi-random superposition of chemical and Knight shifts. As quadrupolar nuclei, it is possible for a second-order quadrupole contribution to affect the mean shift, however this contribution should be small. For example, Nb oxides having a nearly symmetric first-neighbor shell are found to have $^{93}$Nb quadrupole parameters $\nu_Q$ less than 1 MHz \cite{papulovskiy2013theoretical}, which would yield \cite{carter1977metallic} a mean shift contribution of about 50 ppm. This is 2\% of the mean shift for the Ti(0.3) sample, and is presumably an upper limit, so we neglect such contributions in analyzing the relative shifts.

\subsubsection{Shifts}

\begin{figure*}
\includegraphics[width=2\columnwidth]{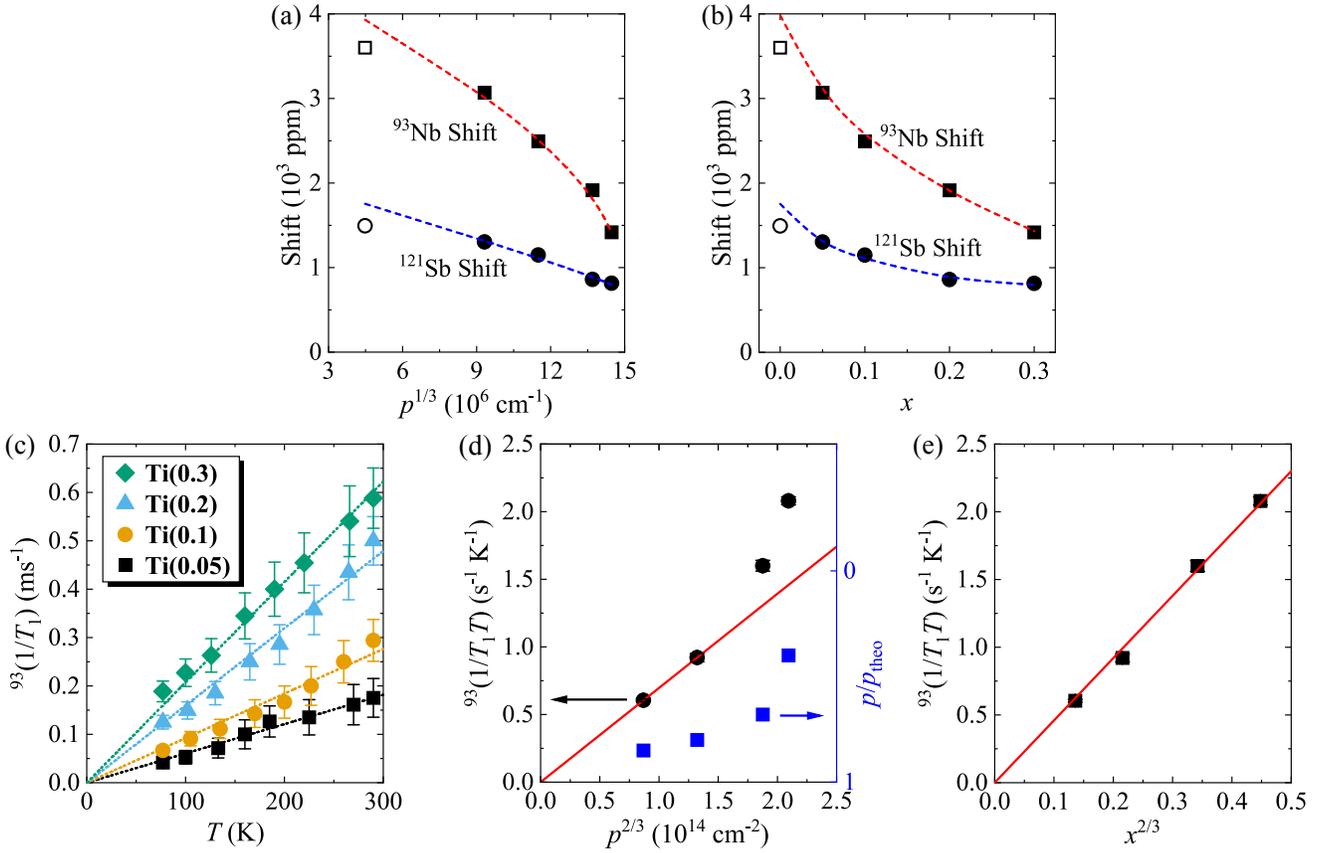}
\caption{\label{Nb1-xTixFeSb_T1andshift}(a) Center-of-mass shifts of $^{93}$Nb and $^{121}$Sb vs $p^{1/3}$ at room temperature, where $p$ is the measured hole density. Dashed lines are the fits to the experimental data described in the text. Open symbols: NbFeSb-1050 sample. (b) The same data and fits as (a) but plotted vs $x$, where $x$ is the Ti content. (c) Temperature dependence of relaxation rates for $^{93}$Nb in Ti($x$) samples, indicated by squares, circles, triangles, and diamonds for $x = 0.05, 0.1, 0.2, 0.3$ respectively, with fits to metallic behavior as described in the text. (d) $^{93}(1/T_1T)$ vs $p^{2/3}$ with the straight line corresponding to a simple filling of a parabolic band, fitted to the two lowest points. Data from the linear fits shown in (c). Also shown: $p/p_\mathrm{theo}$ vs $p^{2/3}$. (e) $^{93}(1/T_1T)$ vs $x^{2/3}$ with a linear fit.}
\end{figure*}

\begin{figure*}
\includegraphics[width=2\columnwidth]{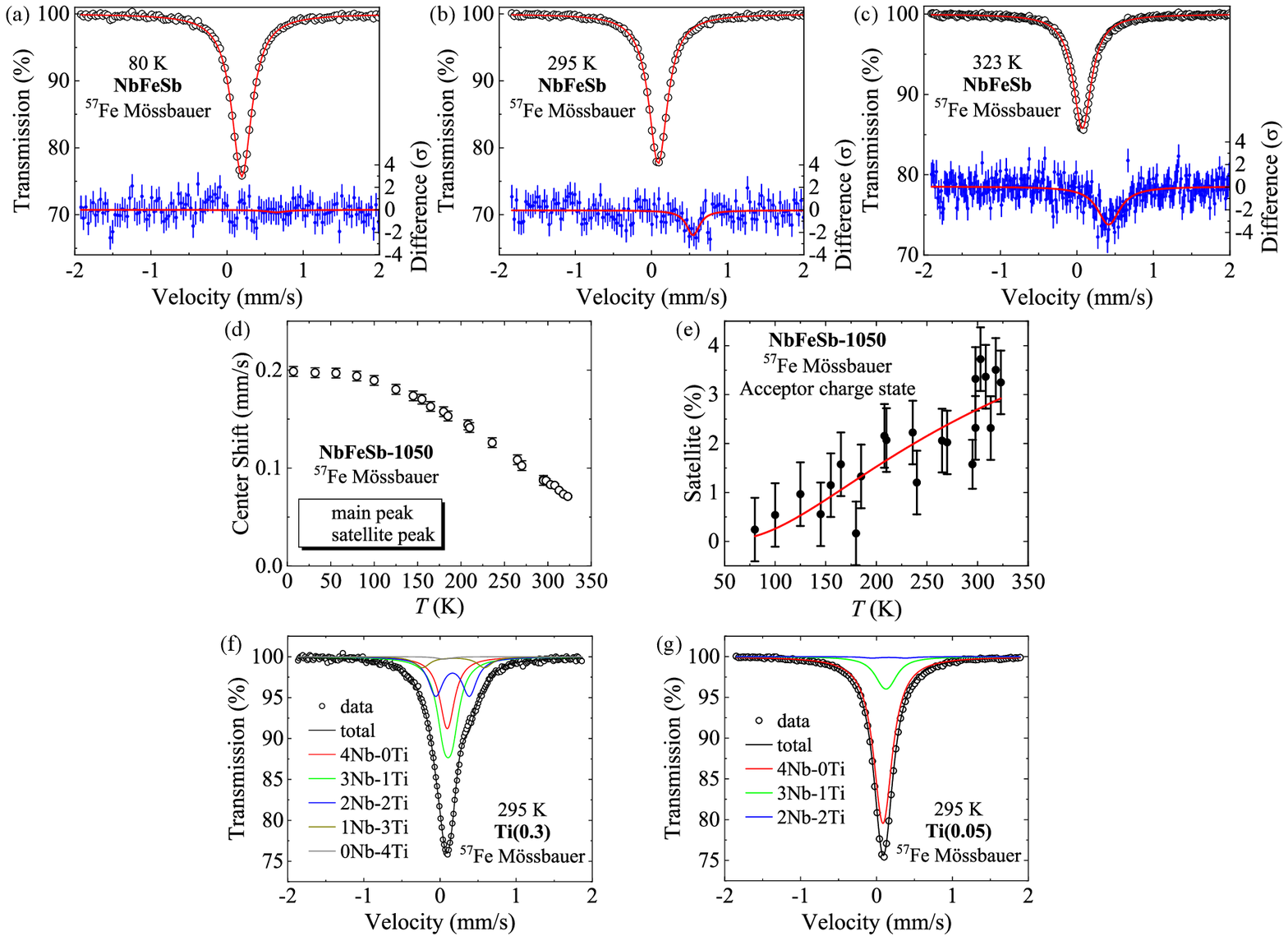}
\caption{\label{Nb1-xTixFeSb_mossbauer} (a)-(c) $^{57}$Fe M\"{o}ssbauer spectra for unsubstituted NbFeSb-1050 sample with least-squares fits described in the text plotted as solid curves. Velocities are relative to $\alpha$-Fe, with error bars too small to be seen. Residuals are also shown separately with fitted satellite curve, relative to the statistical error of the fit. (d) Shift vs temperature for the fitted NbFeSb-1050 main peak. (e) Satellite peak relative area vs temperature. (f)-(g) $^{57}$Fe M\"{o}ssbauer spectra of Ti(0.3) and Ti(0.05) samples, with fits for neighbor configurations of Fe atoms as shown. 3Nb-1Ti and 4Nb-0Ti configurations have negligible probabilities for Ti(0.05) and are not shown.}
\end{figure*}

$^{93}$Nb and $^{121}$Sb shifts for all samples, measured at room temperature, are shown in Fig.~\ref{Nb1-xTixFeSb_T1andshift}(a). These are the center-of-mass isotropic positions of the measured spectra defined as the intensity-weighted average shifts, thus corresponding to the mean shift of the observed nuclei. These shifts can be considered as a sum of the Knight shift ($K$) due to the susceptibility of the charge-carrier spins and the chemical shift ($\delta$) due to the local orbital susceptibility. For samples sufficiently doped to exhibit metallic behavior, $K$ is given generally as,
\begin{equation}
K=\frac{H_\mathrm{HF}\chi_P}{\mu_B},
\end{equation}
where $H_\mathrm{HF}$ is the relevant hyperfine coupling field constant, $\mu_B$ is Bohr magneton, and $\chi_P$ is the Pauli electron spin susceptibility per atom, $g(E_F)(g_\mathrm{eff}/2)\mu_B^2$ for weakly-interacting electrons. $g_\mathrm{eff}$ is the effective $g$-factor due to spin-orbit coupling, which can modify the energy splitting and also $K$. For $s$-character conduction electron states, the dominant hyperfine interaction is Fermi contact. However, with $d$ electrons dominant here, the core polarization hyperfine field $H_\mathrm{CP}$ is the relevant spin coupling with the dipolar spin contribution to $K$ vanishing in cubic symmetry. Note also that we assume spin-orbit coupling effects are small. As a result, $K$ can be expressed as
\begin{equation} \label{knight_shift}
K=g_\mathrm{partial}(E_F)(\frac{g_\mathrm{eff}}{2})\mu_BH_\mathrm{CP},
\end{equation}
where $g_\mathrm{partial}(E_F)$ is the Fermi-level partial density of states for the atom containing the nucleus being measured. 

In an effective mass approximation, which is often appropriate for semiconductors, it is found in the metallic limit,
\begin{equation} \label{dos}
g(E_F)=m_\mathrm{eff}\frac{(3\pi^2n)^{1/3}}{\pi^2\hbar^2}V_\mathrm{f.u.},
\end{equation}
where $m_\mathrm{eff}$ is the thermodynamic effective mass, $n$ is the carrier density and $V_\mathrm{f.u.}$ is the volume per formula unit. Thus substituting Eq.~(\ref{dos}) into Eq.~(\ref{knight_shift}), $K$ should scale as $p^{1/3}$. As shown in Ref.~\cite{beshah1987te}, the chemical shift ($\delta$) of Cd$_{1-x}$Zn$_x$Te is linearly dependent on $x$, indicating a linear relationship between substitution fraction and chemical shift similar to other properties often observed in semiconductor alloys. $\delta(x)$ is thus assumed to be linearly dependent on the substitution fraction of Ti. We therefore model the shift as,
\begin{equation} \label{shift}
\begin{split}
(\mathrm{Total~Shift}) & =K(p)+\delta(x) \\
& =A\cdot p(x)^{1/3}+B\cdot x+C
\end{split}
\end{equation}
where $A=\frac{(3\pi^2)^{1/3}}{\pi^2\hbar^2}(\frac{g_\mathrm{eff}}{2})m_\mathrm{eff}\mu_BH_\mathrm{CP}$ and $C$ represents the baseline shift, corresponding to the Fermi level at the mid-gap for pure NbFeSb. The hole densities $p(x)$ are given in Table~\ref{table1}. Curves shown in Fig.~\ref{Nb1-xTixFeSb_T1andshift}(a) and (b) correspond to the fits to Eq.~(\ref{shift}). This yields, for Nb $A=(-1.6 \pm 0.5)\times10^{-4}$ cm, $B=-3150 \pm 1170$ and $C=4700 \pm 500$; and for Sb $A=(-0.9 \pm 0.4)\times10^{-4}$ cm, $B=-200 \pm 900$ and $C=2180 \pm 400$, all in ppm shift units. For all fitting parameters, $B$ has the largest standard error due to the lack of points for $x$ close to 1.

While the $T_1$ results described below allow a more targeted analysis of the carrier behavior, the shift analysis is particularly important in assessing the NbFeSb ($x=0$) chemical shifts. The fitted results indicate a large $^{93}$Nb chemical shift of $\delta\approx4700$ ppm for NbFeSb, decreasing rapidly vs $x$. The $x = 0$ result is slightly larger than previously obtained \cite{tian2018native} since here we are able to better separate the chemical shift. The $^{93}$Nb shift in NbFeSb is quite large; the calculated chemical shift, with the offset for the reference standard calibrated against LaNbO$_4$ and YNbO$_4$, is $\delta=3220$ ppm, a considerably smaller value. We also made a similar calculation for the half-Heusler compound NbCoSn, which yielded a $^{93}$Nb chemical shift of 2585 ppm, in much closer agreement with the measured value, $\delta= 2849$ ppm \cite{tian2020}. The fitted results also indicate a much smaller reduction in $^{121}$Sb chemical shift going from NbFeSb to TiFeSb. The $^{121}$Sb shifts calculated using the WIEN2k package decrease by 132 ppm, following the same trend vs substitution as the measured results. However, the $^{93}$Nb result exceeds the usual range of reported shifts, and here we see that it is significantly larger than obtained by DFT.

\begin{table*}
\caption{\label{table2} Absorption areas discussed in text, and fitted shift parameters for $^{57}$Fe M\"{o}ssbauer spectra of Ti(0.3) and Ti(0.05) samples.}
{\footnotesize
\begin{tabular}{>{\centering}p{1.5cm}>{\centering}p{1.8cm}>{\centering}p{1.8cm}>{\centering}p{1.8cm}>{\centering}p{1.8cm}>{\centering}p{1.8cm}>{\centering}p{1.8cm}}
\hline\hline
\addlinespace[1ex] & \multicolumn{2}{c}{Absorption area (\%)} & \multicolumn{2}{c}{Isomer shift (mm/s)} & \multicolumn{2}{c}{Quadruple splitting (mm/s)} \tabularnewline[1ex] \hline
\addlinespace[1ex] & Ti(0.3) & Ti(0.05) & Ti(0.3) & Ti(0.05) & Ti(0.3) & Ti(0.05) \tabularnewline[0.5ex]\hline
\addlinespace[1ex]4Nb-0Ti & 26.5 & 81.4 & 0.087 & 0.085 & 0 & 0 \tabularnewline[0.5ex]
\addlinespace[1ex]3Nb-1Ti & 44.8 & 17.1 & 0.111 & 0.125  & 0.105 & 0 \tabularnewline[0.5ex]
\addlinespace[1ex]2Nb-2Ti & 19.8 & 1.3 & 0.173 & 0.164 & 0.446 & 0 \tabularnewline[0.5ex]
\addlinespace[1ex]1Nb-3Ti & 8.0 & - & 0.181 & - & 0.725 & - \tabularnewline[0.5ex]
\addlinespace[1ex]0Nb-4Ti & 0.9 & - & $-0.050$ & -  & 0 & - \tabularnewline[0.5ex]
\hline\hline
\end{tabular}
}
\end{table*}

\subsubsection{Spin-lattice relaxation rates}

The $^{93}$Nb spin-lattice relaxation rate, denoted as $^{93}(1/T_1)$, was measured using the inversion recovery method from 77-290 K. The recovery of the $^{93}$Nb central-transition magnetization can be expressed as
\begin{multline}
\frac{M(t)-M(\infty)}{M(\infty)}=-2\alpha(0.152e^{-\frac{t}{T_1}}+0.14e^{-\frac{6t}{T_1}}\\
+0.153e^{-\frac{15t}{T_1}}+0.192e^{-\frac{28t}{T_1}}+0.363e^{-\frac{45t}{T_1}}).
\end{multline}
Here, $M(t)$ is the measured signal and $t$ the recovery time. Each experimental $T_1$ value was obtained by a fit to this multi-exponential recovery curve. For all studied compounds, the $^{93}(1/T_1)$ results exhibit a constant $T_1T$ behavior within error bars, indicating a metallic-type relaxation process as shown in Fig.~\ref{Nb1-xTixFeSb_T1andshift}(c). In the case that an effective mass treatment is appropriate, from an analysis similar to what is given above for $K$ one finds that $1/T_1$ should scale as $p^{2/3}$. As shown in Fig.~\ref{Nb1-xTixFeSb_T1andshift}(d), $^{93}(1/T_1T)$ follows a linear dependence on $p^{2/3}$ for smaller $p$, although there is an enhancement for large $p$, which is further discussed below. On the other hand, with $p$ expected to be proportional to $x$ in the case that each Ti donates one hole to the valence band, we find that indeed the fitted $1/T_1T$ is proportional to $x^{2/3}$, as shown in Fig.~\ref{Nb1-xTixFeSb_T1andshift}(e). This is an indication that the Hall results do not represent all of the holes in the valence band for large substitution levels.

\subsection{M\"{o}ssbauer measurements}

Figs.~\ref{Nb1-xTixFeSb_mossbauer}(a)-\ref{Nb1-xTixFeSb_mossbauer}(c) show M\"{o}ssbauer spectra for the unsubstituted NbFeSb-1050 sample at 80, 295 and 323 K, respectively. The spectra show no sign of magnetic splitting. Least-square fitting curves are also shown in the figure. The initial fits including one singlet revealed a second satellite peak, especially at larger $T$, shown in the residual plots in Figs.~\ref{Nb1-xTixFeSb_mossbauer}(a)-\ref{Nb1-xTixFeSb_mossbauer}(c). Thus we adopted a fitting with two Lorentzian lines, including a main peak and a small satellite with about 0.5 mm/s larger shift. The fitted line widths for the main peak are nearly temperature independent, FWHM $0.30$ mm/s for the main line at 295 K, increasing to $0.32$ mm/s at 80 K. These small line widths indicate a lack of inhomogeneous broadening, showing NbFeSb to be well-ordered in the half-Heusler structure. The main line shift is 0.089 mm/s at 295 K, gradually increasing to 0.199 mm/s at 7 K, very similar to the results reported \cite{hobbis2019structural} for $n$-type NbFeSb. The $T$-dependent shift is shown in Fig.~\ref{Nb1-xTixFeSb_mossbauer}(d). The fitted satellite position is $0.545\pm 0.036$ mm/s at 295 K. Due to the small amplitude of the satellite peak, it was fitted with the same width as the main peak.

The change in amplitude of the fitted satellite vs $T$ can be understood as a change in the charge state of a native defect vs temperature. Fig.~\ref{Nb1-xTixFeSb_mossbauer}(e) shows the spectral area of the satellite peak, as a percentage of the total area, corresponding to the relative number of Fe ions affected by these defects. Since the defect identity is not determined \cite{tian2018native}, the satellite might correspond to either a Fe-centered defect, or immediate Fe neighbors of the defect site, although the lack of quadrupole splitting tends to indicate the former. With a large density of such defects close to the band edge energy, when temperature increases the charge state of these defects can change, as holes are transferred to the valence band, due to excitation of carriers out of the defect level. Accordingly, the result was fit to the acceptor density function \cite{ashcroft1976solid},
\begin{equation}
N_a^+=\frac{N_A}{1+4\cdot e^{\Delta/kT}},
\end{equation}
assuming each neutral acceptor contains two electrons with opposite spins and the state with no electron is prohibited, and where $\Delta$ is the energy difference between defect level and chemical potential. The fitting [Fig.~\ref{Nb1-xTixFeSb_mossbauer}(e)] gives $\Delta=31\pm0.6$ meV and $N_A=39\pm10\,(\%)$ showing maximally around 8\% Fe affected by defects. This is in good agreement with the result previously obtained from NMR \cite{tian2018native}, in which the spin-lattice relaxation rate of NbFeSb also shows a sharp increase close to room temperature, modeled as a large density of shallow defects located about 30 meV above the valence band in the $p$-type samples making up an impurity band near the band edge.

Figs.~\ref{Nb1-xTixFeSb_mossbauer}(f) and \ref{Nb1-xTixFeSb_mossbauer}(g) show $^{57}$Fe M\"{o}ssbauer spectra of Ti(0.3) and Ti(0.05) respectively at room temperature. For Fe atoms ($4c$ sites) in Nb$_{1-x}$Ti$_x$FeSb alloys, 4 out of 8 nearest-neighbor sites are occupied by mixed Nb and Ti atoms. Therefore, the resulting spectra were modeled as superpositions of peaks for different local configurations of these neighbor ions. The spectra were fit assuming amplitudes corresponding to a binomial distribution,
\begin{equation}\label{binomial}
P_n(x)=\frac{4!}{n!(4-n)!}x^n(1-x)^{4-n},
\end{equation}
for the probability of each Nb$_{4-n}$Ti$_n$ configuration. For the Ti(0.3) composition, probabilities given by Eq.~(\ref{binomial}) are 24, 41, 26, 7.6, and 0.8\%, for $n=0$ through 4 respectively. However, after first fixing these areas in the fit, we found that the goodness of fit ($\chi^2$) decreased from 1.58 to 0.98 based on slightly modified probabilities equal to 26.5, 44.8, 19.8, 8.0, and 0.9\%, for the same neighbor configurations. The results are plotted in Fig.~\ref{Nb1-xTixFeSb_mossbauer}(f) based on the modified probabilities. The fitting parameters are shown in Table~\ref{table2}; widths were held fixed in these fits. The results indicate a reduced probability for the 2Nb-2Ti configuration in favor of the others, which could be an indication of the segregation of Nb and Ti neighbors for larger substitution amounts. On the other hand, for Ti(0.05), the standard binomial probabilities, 81.4, 17.1, and 1.3\% (Table~\ref{table2}), with negligible contributions for 3 and 4 neighbors, worked very well in fitting the data. In the fits, the peaks for Fe with Nb-only neighbors (4Nb-0Ti) have very similar center shifts of 0.087 and 0.085 mm/s, and these are very close to the room-temperature shifts for the unsubstituted sample with an identical Fe nearest neighbor configuration. The increasing isomer shift vs number of Ti neighbors, shown in the table, is an indication of enhanced $d$-electron transfer to Fe in these configurations.

\subsection{Magnetic measurements}

\begin{figure}
\includegraphics[width=0.8\columnwidth]{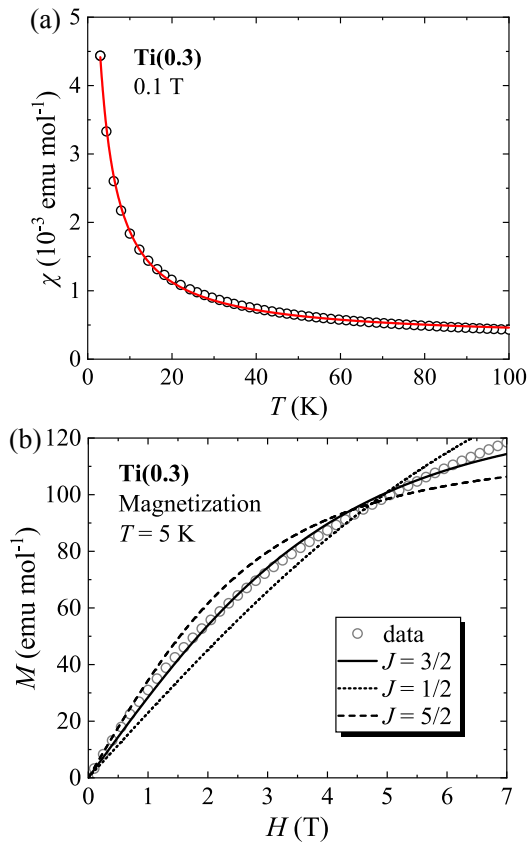}
\caption{\label{Nb1-xTixFeSb_magnetism} Magnetic measurements for sample Ti(0.3). (a) Susceptibility vs $T$ from 3 to 100 K with fit described in text. (b) $M$ vs $H$ measured at 5 K. $J=3/2$ Brillouin function fit is shown with $J = 1/2$ and $J = 5/2$ curves for comparison.}
\end{figure}

The magnetic susceptibility ($\chi$) of the sample with the largest Ti concentration, Ti(0.3), is shown in Fig.~\ref{Nb1-xTixFeSb_magnetism}(a), for a fixed field of 1000 Oe. We fit the low-$T$ data to a Curie-Weiss function according to the standard relationship, $\chi(T)=C/(T-T_c)+\chi_0$, where $C=N_Acp_\mathrm{eff}^2\mu_B^2/3k_B$ is the Curie constant with $c$ the concentration of paramagnetic defects, and $p_\mathrm{eff}$ the effective moment. The results are $T_C=-1.4$ K, $\chi_0=2.8\times10^{-4}$ emu\,mol$^{-1}$ and assuming $p_\mathrm{eff} = 3.87$ corresponding to $J=3/2$ (see below), a dilute concentration $c = 0.01$ per formula unit of these defects.

$M$ vs $H$ measurements confirm that the magnetic response is due to dilute paramagnetic defects, as shown at $T = 5$ K in Fig.~\ref{Nb1-xTixFeSb_magnetism}(b) for sample Ti(0.3). To analyze for the local magnetic moments, data were fit to
\begin{equation}
M = N_AcgJ\mu_BB_J(x),
\label{paramagnetization}
\end{equation}
where $B_J(x)$ is a Brillouin function with $x = \frac{g\mu_BJB}{k_BT}$. Assuming $g=2$ expected for transition ions, we found that $J = 3/2$ gives the closest agreement [Fig.~\ref{Nb1-xTixFeSb_magnetism}(b)] by choosing possible $J$ values and fitting to $c$. Fixing $J=3/2$, the fitted $c=0.008$ per formula unit agrees well with the value obtained from $M$-$T$ measurement, $c=0.01$. This indicates that the predominant magnetic defect is a $J=3/2$ local moment.

The small density of moments obtained here is comparable to what was obtained in annealed unsubstituted NbFeSb samples \cite{tian2018native}, for example 0.002 per formula unit in sample NbFeSb-1050 also with $J=3/2$. Thus we see that there is almost no tendency for Ti substitution to promote magnetic defect formation. Note also that in TiFe$_{1+x}$Sb half-Heusler alloys close to the stable TiFe$_{1.33}$Sb composition \cite{tobola2001composition,tavassoli2018half}, with the Fe interstitials balancing the charge of the Ti ions to achieve 18-electron balance, a large Curie-type response is observed. However, this term corresponds to only a few percent of the Fe sites, with the bulk of the material also nonmagnetic similar to NbFeSb, as confirmed by Mössbauer results \cite{tavassoli2018half}. In the compositions studied here, even with Ti substitution far from the 18-electron stability rule, there are not significant numbers of Fe interstitials (Table~\ref{table1}). By analogy with the TiFe$_{1.33}$Sb results quoted above, those antisites present are likely to be nonmagnetic, and perhaps these are the source of the defects indicated by M\"{o}ssbauer results. As before \cite{tian2018native} a likely candidate for the dilute-magnetic $J=3/2$ defects would be Fe antisites on Nb positions. Note also that a small paramagnetic peak \cite{tian2018native} might also be expected in the NMR $1/T_1$ [Fig.~\ref{Nb1-xTixFeSb_T1andshift}(c)], however here the relaxation rate is strongly dominated by a larger $1/T_1$ contribution due to the carriers in the substituted samples. 

\section{Discussion}

The variation in amplitude of the satellite $^{57}$Fe M\"{o}ssbauer peak in NbFeSb is an unusual feature corresponding to a relatively large density of defects with charge states changing vs $T$. As shown above, this is consistent with the existence of the impurity band shown previously from NMR results \cite{tian2018native}. The temperature dependence of $^{57}$Fe M\"{o}ssbauer absorption area shows that the origin of the impurity band is Fe-related, perhaps due to Fe interstitials \cite{tian2018native}.

Besides our work, Hobbis $et$ $al.$ \cite{hobbis2019structural} also reported $^{57}$Fe M\"{o}ssbauer spectroscopy of NbFeSb, however in contrast with the $p$-type NbFeSb-1050 sample, their work focused on $n$-type NbFeSb. For $n$-type NbFeSb, two extra small doublets were observed with positions that differ from that of our measured satellite (0.65 mm/s). These different behaviors indicate distinct types of defects, which is reasonable given the different carrier types. 

Comparing to $^{57}$Fe M\"{o}ssbauer results for different materials within the half-Heusler family, for VFeSb, only a single line was observed with an isomer shift of $\sim$0.05 mm/s with no extra peaks \cite{jodin2004effect}. In Ref.~\cite{tavassoli2018half}, $^{57}$Fe M\"{o}ssbauer of TiFe$_{1.33}$Sb also shows two sets of peaks, a doublet (main peak) with an isomer shift 0.108 mm/s and a singlet with an isomer shift 0.279 mm/s (a satellite peak). These were matched to certain local atomic arrangement close to Fe atoms, indicating non-randomness of Fe atoms on the $4d$ site. The larger shifts observed for NbFeSb are consistent with the results found here for Nb$_{1-x}$Ti$_x$FeSb, with an isomer shift which is enhanced as the number of Ti neighbors increases.

In transition metals, besides the contribution from the spin moments of the conduction electrons, the orbital contribution to the NMR relaxation caused by fluctuating orbital moments of the conduction electrons can also make an important contribution to $1/T_1$. In this case, the spin-lattice relaxation rate should be dominated by two terms: $(1/T_1)_\mathrm{total}=(1/T_1)_\mathrm{orb}+(1/T_1)_d$, where the first term is an orbital contribution term and the second is the $d$-spin relaxation rate.

For transition metals with cubic structure, the orbital relaxation rate can be expressed in a general form \cite{obata1963nuclear},
\begin{equation} \label{orbital}
(1/T_1)_\mathrm{orb}=2A\frac{2\pi}{\hbar}[\gamma_e\gamma_n\hbar^2g_\mathrm{Nb}(E_F)\langle r^{-3} \rangle]^2k_BT,
\end{equation}
where $A=10C(2-C)$ is a dimensionless quantity with $C$ the degree of admixture of $\Gamma_5$ and $\Gamma_3$ symmetry at the Fermi level and $\langle r^{-3} \rangle$ is the average over occupied $d$ orbitals, expected \cite{knigavko2007divergence} for NbFeSb to include only on-site contributions. It was shown in Ref.~\cite{obata1963nuclear} that $(1/T_1)_\mathrm{orb}$ for $d$-band metals reaches a maximum with an admixture of $d$ orbitals corresponding to $\Gamma_5 : \Gamma_3 = 3:2$. Our DFT calculation shows that the ratio of atomic functions for NbFeSb near the band edge is $t_{2g} (\Gamma_5):e_g (\Gamma_3) = 68\%:32\%$, giving $A=9.8$, close to the maximum $A=10$, while for Nb, the calculated $g(E)$ near the VB edge is 14\% of the total. Since, as shown below, $(1/T_1)_\mathrm{orb}$ is found to dominate, we examine the dependence on Ti substitution. In Fig.~\ref{Nb1-xTixFeSb_T1andshift}(d), the solid line is a fit of Eq.~(\ref{orbital}) to Ti(0.05) and Ti(0.1) with $g(E_F)$ expressed by Eq.~(\ref{dos}). Using $\langle r^{-3} \rangle=1.84\times10^{25}$ cm$^{-3}$ \cite{koh1985hyperfine}, we obtain $m_\mathrm{eff}=4.6\,m_e$ in a good agreement with the calculated $m_\mathrm{eff}=4.9\,m_e$ for unsubstituted NbFeSb.

The core polarization contribution to the $d$-spin spin-lattice relaxation rate in metals is
\begin{equation} \label{spin}
(1/T_1)_d=2hk_BT[\gamma_nH_\mathrm{CP}g_\mathrm{Nb}(E_F)]^2q,
\end{equation}
where the core polarization hyperfine field $H_\mathrm{CP}$ is reported to be $-21$ T \cite{yafet1964nuclear} and $q$ is a reduction factor which is a function of the admixture of $d$ orbitals. In the present case, nearly uniform occupation of the five $d$ orbitals gives $q\approx1/5$ \cite{yafet1964nuclear}. Using $g_\mathrm{Nb}(E_F)=0.243$ states/eV calculated as described above, Eq.~(\ref{spin}) gives $(1/T_1T)_d=0.016$ s$^{-1}$\,K$^{-1}$ with $m_\mathrm{eff}=4.6\,m_e$, inserted in Eq.~(\ref{dos}). This is considerably smaller than the observed rates. There is also a dipole spin contribution to $1/T_1$, however for the large $d$-orbital degeneracy case, this can be shown \cite{yafet1964nuclear} to be much smaller than the orbital contribution. These results show that the orbital contribution is the dominant term in the spin-lattice relaxation process.

When the Ti fraction increases to 30\%, the $1/T_1T$ values depart from $p^{2/3}$ behavior (solid line in Fig.~\ref{Nb1-xTixFeSb_T1andshift}(d) fitted to the two low-$p$ points). At the same time, there is a decrease in the measured hole density relative to $p_\mathrm{theo}$, as also shown in Fig.~\ref{Nb1-xTixFeSb_T1andshift}(d). Non-parabolicity of the valence band could give an increase in $m_\mathrm{eff}$ which might explain the $1/T_1T$ upturn, however such an effect would not be expected to affect the Hall effect results, at least in the spherical hole pocket limit. Thus, we conclude that the presence of resonant levels having low mobility in the valence band \cite{heremans2012resonant} becomes important for large $x$. The results shown in Fig.~\ref{Nb1-xTixFeSb_T1andshift}(e) help to further clarify this result, indicating that $1/T_1$ is affected by states near $E_F$ which make little contribution to the Hall results. For large $x$, the Fermi level moves more deeply into the valence band encountering states caused by Ti substitutions. This effect could have significance for thermoelectric properties, however, the agreement between theory and experiment indicates that such effects are not important for smaller $x$, and thus a rigid-band effective mass model provides a good description for the less-heavily substituted compositions.

For Ti($x$) samples, the measured Knight shift values are also found to be larger than expected. For example, the Knight shift contribution to the total shift for the Ti(0.3) sample is $-2185$ ppm from our fit. However, using $g_\mathrm{Nb}(E_F)=0.243$ states/eV obtained above and the core polarization hyperfine field $H_\mathrm{CP}=-21$ T, Eq.~(\ref{knight_shift}) gives $K=-295$ ppm. The Knight shift difference could be explained by a large $g_\mathrm{eff}$ in these samples, or by electron-electron interactions which can also enhance the measured spin susceptibility and thereby the Knight shift.

In addition to the observed large Knight shifts, the $^{93}$Nb chemical shifts ($\delta$) also show relatively large values as noted above. The DFT calculations give the $^{93}$Nb chemical shift as 3268 ppm, compared to the fitted $\delta=4700$ ppm. The difference corresponds to an enhancement of the local orbital hyperfine field which is related to a Van Vleck susceptibility ($\chi_\mathrm{VV}$). In addition to the well-known enhancement of the spin susceptibility, electron-electron interactions can also lead to an enhancement of $\chi_\mathrm{VV}$ \cite{kontani1996magnetic,kontani1997electronic}, thus yielding a large chemical shift. Ref.~\cite{kontani1997electronic} also pointed out that orbital degeneracy is necessary for this effect; the large degeneracy of orbitals was already discussed above in analyzing the $1/T_1T$ of these samples. Thus, although the calculated distribution of $d$ electrons in the valence band for Nb and Fe provide good agreement with the observed orbital $T_1$, as a measure of the local susceptibilities the NMR shifts demonstrate that electron-electron interaction effects are relatively strong for these states.

\section{Conclusions}

In this work, we have investigated the effect of defects and substitutions of NbFeSb and Nb$_{1-x}$Ti$_x$FeSb using local NMR and M\"{o}ssbauer probes. The spin-lattice relaxation results are well modeled in terms of an orbital contribution in good agreement with DFT calculations for NbFeSb. With increasing $x$, we find a deviation from the expected behavior which we understand as due to resonant valence band levels which do not contribute to transport results. NMR shift vs $x$ are well-explained by a model combining carrier-concentration-dependent Knight shift and composition-dependent chemical shift. The local paramagnetic susceptibilities are found to be significantly enhanced relative to calculated values. The $T$-dependence of the satellite peak in the unsubstituted NbFeSb M\"{o}ssbauer spectrum provides a direct measure of charging of acceptor states in an impurity band located around 30 meV above valence band edge as previously reported. The M\"{o}ssbauer spectrum for $x=0.3$ shows small departures from a binomial distribution, indicating a small deviation from random substitution in the mixed alloys, revealing a possible segregation of 4$a$-site substitution atoms.

\begin{acknowledgments}
This work is supported by the Robert A. Welch Foundation, Grant No. A-1526. F. G. V. also acknowledges the support by National Science Foundation, Grant No. PHY-150-64-67.
\end{acknowledgments}

\bibliography{Nb1-xTixFeSb}

\end{document}